# Effect of Gd/Nd doping on the magnetic properties of PrMnO$_3$

**Sanjay Biswas and Sudipta Pal**[*]

Department of Physics, University of Kalyani, Kalyani, Nadia, W.B., 741235, India

[*]Author for correspondence: Sudipta Pal, e-mail: sudipta.pal@rediffmail.com

Tel No. : 091 033 2582 2505

Fax      : 091 033 2582 8282

**Abstract**

A study on temperature dependent magnetic properties of single phase orthorhombic perovskites system associated with space group *Pbnm* compounds Pr$_{1-x}$(Gd/Nd)$_x$MnO$_3$ ($x$=0.3, 0.5, 0.7) was carried out. A magnetization reversal is observed below the Neel temperature ($T_N$), in DC magnetization measurements (at 50 Oe) in the doped compounds. This may be due to the antiparallel coupling between the two magnetic sublattices (|Pr+Gd/Nd | and Mn). With lowering of temperature, the |Pr+ Gd/Nd|) ions begin to polarize under the negative internal field due to canted moment of Mn moments. The hysteresis plot taken at 50K shows a ferrimagnetic characteristic and the presence of spin canting of ions in the magnetic sublattices. Arrott plot indicates field induced second order paramagnetic to ferrimagnetic (PM-FiM) phase transition in this system.

## 1. Introduction

Rare earth transition metal oxides ($RTmO_3$) are still centre of attention due to their potential application as multifunctional materials [1–6]. A large number of studies have been done since the last decade to elucidate the nature of spin, charge, orbital ordering or the exchange interaction of the transition metals. The magnetic moments of the rare-earth ions have been polarized due to the coupling with the Mn subsystem and affect the low-temperature magnetic and thermodynamic properties of the manganites as observed in $PrMnO_3$ and $NdMnO_3$ [6, 7]. Yoshii *et al* [8] studied $GdCrO_3$ and showed that rare-earth ions (Gd) and transition metal ions (Cr) canted or aligned in opposite direction. In $Nd_{1-x}Ce_xMnO_3$ manganite the low temperature magnetization reversal phenomena described by the ferrimagnetic coupling between $Mn^{2+}/Mn^{3+}$ and $Mn^{3+}/Mn^{3+}$ sublattices [9]. It has been shown earlier [5] that in $Gd_{0.7}Ca_{0.3}Mn_{1-x}Cr_xO_3$, Gd ions aligned against the Cr and Mn ions. N Sharma *et al* [10] showed that for being $La^{3+}$ non-magnetic, reversal of magnetization in $La_{0.5}Gd_{0.5}CrO_3$ is due to canted opposite direction magnetic moment of Gd and Cr. Again in $Sr_2YbRuO_6$ [11] Ru and Yb are canted in opposite direction. This type of reversal of magnetization phenomena has also been discovered in $La_{1-x}Pr_xCrO_3$ [12], $Bi_{0.3}Ca_{0.7}Mn_{0.75}Cr_{0.25}O_3$ [3], $La_{1-x}Gd_xMnO_3$ [13], $La_{1-x}Ce_xCrO_3$ [14] and explained considering two sublattices system which are aligned antiparallel to each other. Again in $YFe_{1-x}Mn_xO_3$ [15] manganites the magnetization reversal has been originated from the random occupation of magnetic ions and different magnetic interactions like Fe-O-Fe, Fe-O-Mn and Mn-O-Mn. In $LaVO_3$ [16], orthovanadate which undergoes a transition from orthorhombic (space group *Pbnm*) to monoclinic (space group $P2_1/b11$) unit cell symmetry below the Neel temperature ($T_N$) which causes the reversal of magnetization.

So the magnetization properties drastically changes with the replacement of one rare earth ion with another. In this paper, we have investigated the magnetic properties of the solid solutions of $Pr_{1-x}Gd_xMnO_3$ and $Pr_{1-x}Nd_xMnO_3$ ($x$=0.3, 0.5, 0.7) and it is interesting to observe how the magnetic properties change with doping of Gd or Nd ions in place of Pr.

## 2. Experimental

The polycrystalline samples $Pr_{1-x}(Gd/Nd)_xMnO_3$ ($x$=0.3, 0.5, 0.7) were prepared following standard solid state ceramic method using high purity oxides $Pr_6O_{11}$, $Gd_2O_3$, $Nd_2O_3$ and $MnO_2$. Starting materials were mixed in proper stoichiometric ratio, ground and then heated at 900°C for 24 h. The mixed powders thus obtained were reground, pelletized and sintered at 1500°C for 24h and cooled down to room temperature at a rate of 100° C /h. Structural characterization was done by powder X-ray diffraction (XRD) measurements using CuK$_\alpha$ radiation (Philips diffractometer). The Rietveld analysis of the XRD data using FULLPROF program [17] gives the lattice parameters.

Quantum Design magnetometer (SQUID) was used for the magnetic measurement. Temperature dependent magnetization (M-T) data were recorded in the temperature range 5 K-200 K in both zero field cooled (ZFC) and field cooled (FC) modes under externally applied magnetic field 50 Oe. The samples were first cooled down to 5 K in zero field for ZFC measurements, an external magnetic field (50 Oe) was applied and magnetization was recorded in the presence of this field in the warming cycle. Again the samples were cooled under the presence of an external field of the same strength then heat the samples upto 200K and magnetization was recorded in the cooling cycle for FC measurements. Magnetization (M) isotherms were recorded at 50K on $Pr_{0.5}Gd_{0.5}MnO_3$ and $Pr_{0.5}Nd_{0.5}MnO_3$ up to an applied magnetic field (H) of ±70 kOe.

3. **Results and Discussion**

Powder X-ray diffraction (XRD) pattern of $Pr_{0.5}Gd_{0.5}MnO_3$ and $Pr_{0.5}Nd_{0.5}MnO_3$ taken at room temperature shown in figure 1. The Rietveld refinement of XRD peaks were indexed to single phase orthorhombic perovskites system associated with space group *Pbnm* in which the following parameter relation is observed: $b>a>c/\sqrt{2}$ for all the samples. This relation indicates a static Jahn-Teller (*J-T*) distortion [13]. The obtained lattice parameters values are in-between the end compounds $GdMnO_3$, $PrMnO_3$, $NdMnO_3$ reported earlier [1, 6]. With the increase of Gd and Nd concentration, the lattice parameters *a* and *c* decrease where as *b* increases and the volume decreases shown in table 1. The parameter $\varepsilon=(b-a)/(a+b)$ describes the orthorhombic distortion [13]. The distortion and lattice parameters changes with the doping concentration (table 1). The change of distortion and lattice parameters due to substitution of $Pr^{3+}$ ions (ionic radius=1.09Å) by the $Gd^{3+}$ ions (ionic radius=0.938Å) is greater than the $Nd^{3+}$ ions (ionic radius=1.08Å) [18].

The temperature dependent DC magnetization of $Pr_{1-x}Gd_xMnO_3$ and $Pr_{1-x}Nd_xMnO_3$ (*x*=0.3, 0.5, 0.7) under the external applied field 50 Oe in ZFC and FC processes is presented in figures 2(a), (b) and 3(a), (b) respectively. The ZFC magnetization displayed a positive maximum below the Neel temperature for all the compounds. The temperature at which magnetization becomes maximum is considered as $T_{max}$. Ferromagnetic (FM) ordering has been observed in $Pr_{0.7}Gd_{0.3}MnO_3$ and $Pr_{0.7}Nd_{0.3}MnO_3$ but missing in other samples because larger magnetic anisotropy arises with increasing doping concentration ($x = 0.5, 0.7$). With decreasing temperature the ZFC magnetization decreases and reaches zero at compensation temperature ($T_{comp}$) 13, 24, 34K for $Pr_{1-x}Gd_xMnO_3$ ($x=0.3, 0.5, 0.7$) respectively and 18K for $Pr_{0.7}Nd_{0.3}MnO_3$. Below the $T_{comp}$, the ZFC magnetization shows a negative value. The reason for this characteristic may be due to the small magnetic moment of Nd ions compare to the Gd ions and larger magnetic anisotropy arises in the sample due to Gd doping than the Nd doping. So, the Pr and Gd ions are more and more aligned opposite to the Mn moments due to the external field and internal field of the Mn moment. So the ferromagnetic network |Pr+Gd| easily surpasses the magnetic moment of the Mn and show the negative magnetization in $Pr_{1-x}Gd_xMnO_3$ ($x=0.3, 0.5, 0.7$). But the ferromagnetic network |Pr+Nd| for $Pr_{1-x}Nd_xMnO_3$ ($x=0.5, 0.7$) does not exceed the magnetic moment of the Mn and unable to show the negative magnetization. The FC magnetization also increases with the decrease of temperature and shows a positive maximum magnetization ($M_{Max}$) at respective $T_{max}$ below the Neel temperature ($T_N$). With decreasing temperature the FC magnetization decreases and reaches zero at the compensation temperature ($T_{comp}$) 14, 29, 36K for $Pr_{1-x}Gd_xMnO_3$ ($x=0.3, 0.5, 0.7$) and 18, 22K for $Pr_{1-x}Nd_xMnO_3$ ($x=0.3, 0.5$) respectively. Below the $T_{comp}$, the magnetization becomes negative and reaches $M_{Min}$ at temperature 5K.

The reversal of magnetization can simply be explained in these samples considering $M_{Mn}$ is parallel to the applied field but $M_{|Pr+A|}$ (A= Gd/ Nd) are anti-parallel to the field. With dropping of temperature, the disordered |Pr+A| (A= Gd/ Nd) ions begin to polarize against the canted field of $M_{Mn}$ moments. The net magnetization of the system becomes $M=M_{Mn}-M_{|Pr+A|}$ (A= Gd/Nd). When the temperature dropping from $T_{max}$, the value of $M_{|Pr+A|}$ (A= Gd/Nd) rises considerably quicker than $M_{Mn}$. At the compensation temperature ($T_{comp}$), the magnetizations of the two sublattices become identical and net magnetization turn into zero. Below the $T_{comp}$ the net magnetization becomes negative value.

The solid lines in figures 2(a), (b) and 3(a), (b) show that the magnetization ($M$) can well be approximated by [5, 6, 11]

$$M = M_{Mn} + \frac{C_{Gd/Nd}(H_{int} + H_{ext})}{(T - \theta_W)} \quad (1)$$

Where $M_{Mn}$, $H_{int}$, $H_{ext}$, $C_{Gd/Nd}$, $\theta_W$ stand for the canted moment of Mn, internal field due to the canted Mn moment, applied field, Curie constant and Weiss constant respectively. The constraint of this analysis has been made to assume that $M_{Mn}$ and $H_{int}$ are independent of temperature, which is usually true if $T<<T_N$. It would be interesting to compare the parameters obtained from the fitting using the equation (1). The positive value of $M_{Mn}$ is consistent with the result that the magnetization goes through a positive maximum before reaching the compensation point and the values changes from 280 emu/mol to 820 emu/mol in FC mode and 47 emu/mol to 290 emu/mol in ZFC mode for different samples. And $M_{Mn}$ have been used as a fitting parameter and considered to be constant well below the temperature $T_{max}$. The Weiss constant $\theta_W$ shows a negative value (upto -9K). The internal field ($H_{int}$) shows also negative value (upto ~ $-10^4$ Oe) for all the samples. The negative value of the internal field ($H_{int}$) directs that its direction is contrary of both the applied field and the canted Mn moments. Comparable results have also been observed in $Gd_{0.7}Ca_{0.3}Mn_{1-x}Cr_xO_3$ [5], $Bi_{0.3}Ca_{0.7}Mn_{0.75}Cr_{0.25}O_3$ [3], $GdCrO_3$ [8], and $La_{0.5}Gd_{0.5}CrO_3$ [10].

The DC susceptibility ($\chi$) above the transition temperature ($T_{max}$) the curve follows a Curie-Weiss behaviour $\chi = \mu^2_{eff} / [8(T-\Theta)]$ as shown in the figures 4(a) and (b). Positive paramagnetic temperature $\Theta$ is obtained for $Pr_{0.7}Gd_{0.3}MnO_3$ but negative $\Theta$ are obtained for $Pr_{1-x}Gd_xMnO_3$ ($x$=0.5 and 0.7). However for $Pr_{1-x}Nd_xMnO_3$ ($x$=0.3, 0.5, 0.7), all samples shows a positive $\Theta$ value. These outcomes implied that, the samples changes from ferromagnetic background to ferrimagnetic background with doping of Gd but ferromagnetic background remain in the Nd doping samples. Effective magnetic moment ($\mu_{eff}$) per mole for all the samples could be obtained from the fitting using Curie-Weiss law. The value of $\mu_{eff}$ for the samples $Pr_{1-x}Gd_xMnO_3$ ($x$=0.3, 0.5, 0.7) are larger than the samples $Pr_{1-x}Nd_xMnO_3$ ($x$=0.3, 0.5, 0.7) due to the large magnetic moment of Gd ions compare to the Nd ions. The quantum mechanical spins of different magnetic ions to calculate $\mu_{eff}^2$ per mole are: $2^2\mu_B^2[(1-x)\times2/2\times4/2]$ form $Pr^{3+}$; $2^2\mu_B^2(x\times7/2\times9/2)$ form $Gd^{3+}$; $2^2\mu_B^2(x\times3/2\times5/2)$ from $Nd^{3+}$; $2^2\mu_B^2(4/2\times6/2)$ from $Mn^{3+}$. Theoretically from the quantum mechanical spins we calculate $\mu_{eff}$ per mole for $Pr_{1-x}Gd_xMnO_3$ ($x$=0.3, 0.5, 0.7) are 6.96$\mu_B$, 7.71$\mu_B$ and 8.4$\mu_B$ respectively and for $Pr_{1-x}Nd_xMnO_3$ ($x$=0.3, 0.5, 0.7) are 5.839$\mu_B$, 5.958$\mu_B$, 6.074$\mu_B$ respectively. The Curie-Weiss law gives from experimental result $\mu_{eff}$ per mole for $Pr_{1-x}Gd_xMnO_3$ (x=0.3, 0.5, 0.7) are 6.79$\mu_B$, 7.95$\mu_B$, 8.17$\mu_B$ respectively and for $Pr_{1-x}Nd_xMnO_3$ (x=0.3, 0.5, 0.7) are 5.548$\mu_B$, 5.429$\mu_B$, 5.774$\mu_B$ respectively. The experimental and theoretical values deviate slightly. This is very common phenomenon in these type of oxides [4, 19]. This result could be easily understood by considering that some of the ions |Pr+A| (A= Gd/ Nd) in the ferromagnetic network aligned opposite or canted from the direction of Mn ions due to the magnetic anisotropy in the compounds. So the net effective magnetic moment differs from theoretically and experimentally calculated result.

For better understanding the magnetic property, field dependent magnetization measurement has been performed at 50K on $Pr_{0.5}Gd_{0.5}MnO_3$ and $Pr_{0.5}Nd_{0.5}MnO_3$. The measurements have been performed after cooling the sample under zero field and the results have been shown in the figure 5. For $Pr_{0.5}Gd_{0.5}MnO_3$ the M-H curve shows a small hysteresis with coercive field ($H_C$) ~980 Oe and for $Pr_{0.5}Nd_{0.5}MnO_3$ the M-H curve shows very small hysteresis (with coercive field ($H_C$) ~350 Oe). A narrow hysteresis loop and a large slope at high fields in M-H curve, suggesting the coexistence of ferromagnetic and antiferromagnetic phases. The hysteresis loops are symmetric about the origin and specify the lack of the exchange bias field for the samples. It is clear due to the spin canting of ions in the magnetic sublattices a large magnetic anisotropy arises in the samples. So the magnetization does not reach to any saturation value, the magnetization increasing almost linearly upto the magnetic field H= ±40000 Oe in the M-H curve.

Figure 6 shows that Arrott plots of isotherms for $Pr_{0.5}Gd_{0.5}MnO_3$ at different temperatures near to $T_{max}$. One can distinguish the order of the phase transition from the slope of the line. According to previous reports [20, 21] the magnetic transition is second order if all the Arrott plots have positive slope and negative slope essentially indicates that the phase transition is of first order. In our case we get a positive slope for all the plots, so it is clear that the compound shows field induced second order paramagnetic to ferrimagnetic (PM-FiM) phase transition. Similar second order phase transition has also been observed for the $Pr_{0.5}Nd_{0.5}MnO_3$.

4. **Conclusions**

The solid solution of rare-earth-based single phase orthorhombic perovskites system $Pr_{1-x}(Gd/Nd)_xMnO_3$ (x=0.3, 0.5, 0.7) was studies. In all the samples for both the doping element Gd or Nd, an internal field generates due to the Mn sublattice which polarizes the rare-earth ions magnetic moments in the opposite direction, resembling the situation found in ferrimagnetic systems. The experimental magnetization reverses its sign and showed a negative value in ZFC and FC process. The magnetic moments of rare-earth network vary with temperature when the magnetic moment is larger than the Mn moment the resultant leads to a negative magnetization. Due to the large magnetic moment of Gd ions compare to the Nd ions, the samples doping with Gd shows larger magnetic anisotropy and magnetization reversal.

**Acknowledgements**

This work was partially supported by DST-FIST and DST-FAST TRACK project No-SR/FTP/PS-101/2010 Govt. of India.

**Table 1.** The values of the lattice parameters and orthorhombic distortion of $Pr_{1-x}(Gd/Nd)_xMnO_3$ ($x$=0.3, 0.5, 0.7).

| Compounds | $a \pm 10^{-3}$ (Å) | $b \pm 10^{-3}$ (Å) | $c \pm 10^{-3}$ (Å) | $V \pm 10^{-3}$ (Å$^3$) | $\varepsilon=(b-a)/(a+b)$ |
|---|---|---|---|---|---|
| $Pr_{0.7}Gd_{0.3}MnO_3$ | 5.4066 | 5.8294 | 7.5372 | 237.5516 | 0.0376 |
| $Pr_{0.5}Gd_{0.5}MnO_3$ | 5.3795 | 5.8408 | 7.5065 | 235.8616 | 0.0411 |
| $Pr_{0.3}Gd_{0.7}MnO_3$ | 5.3537 | 5.8492 | 7.4780 | 234.1726 | 0.0442 |
| $Pr_{0.7}Nd_{0.3}MnO_3$ | 5.4378 | 5.8163 | 7.5749 | 239.58 | 0.0336 |
| $Pr_{0.5}Nd_{0.5}MnO_3$ | 5.4316 | 5.8195 | 7.5687 | 239.24 | 0.0344 |
| $Pr_{0.3}Nd_{0.7}MnO_3$ | 5.4239 | 5.7977 | 7.5597 | 237.73 | 0.0333 |

**Figure Captions**

**Figure 1.** X-Ray Diffraction pattern of $Pr_{0.5}Nd_{0.5}MnO_3$ and $Pr_{0.5}Gd_{0.5}MnO_3$ observed at room temperature.

**Figure 2.** DC magnetization (M) versus temperature (T) in ZFC mode of **(a)** $Pr_{1-x}Gd_xMnO_3$ ($x$=0.3, 0.5, 0.7) and **(b)** $Pr_{1-x}Nd_xMnO_3$ ($x$=0.3, 0.5, 0.7). Solid line shows fitting of M-T data recorded in the warming cycle (red arrow indicates) under 50 Oe.

**Figure 3.** DC magnetization (M) versus temperature (T) in FC mode of **(a)** $Pr_{1-x}Gd_xMnO_3$ ($x$=0.3, 0.5, 0.7) and **(b)** $Pr_{1-x}Nd_xMnO_3$ ($x$=0.3, 0.5, 0.7). Solid line shows fitting of M-T data recorded in the cooling cycle (red arrow indicates) under 50 Oe.

**Figure 4.** Temperature dependent inverse magnetic susceptibility of **(a)** $Pr_{1-x}Gd_xMnO_3$ ($x$=0.3, 0.5, 0.7) and **(b)** $Pr_{1-x}Nd_xMnO_3$ ($x$=0.3, 0.5, 0.7). Solid line is the high temperature fit using equation $\chi = \mu^2_{eff} / [8(T-\Theta)]$.

**Figure 5.** M-H curves recorded for $Pr_{0.5}Gd_{0.5}MnO_3$ and $Pr_{0.5}Nd_{0.5}MnO_3$ at 50K.

**Figure 6.** Arrott plot of magnetization for $Pr_{0.5}Gd_{0.5}MnO_3$ observed at different temperature.

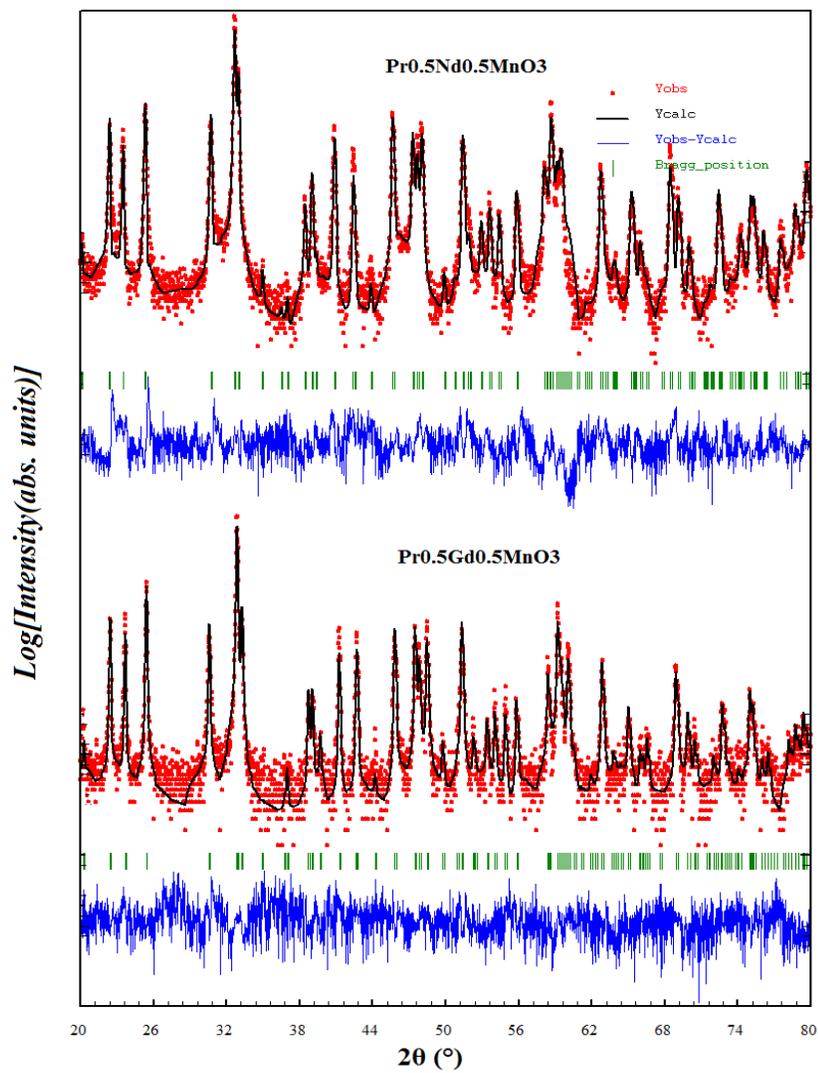

**Figure 1**

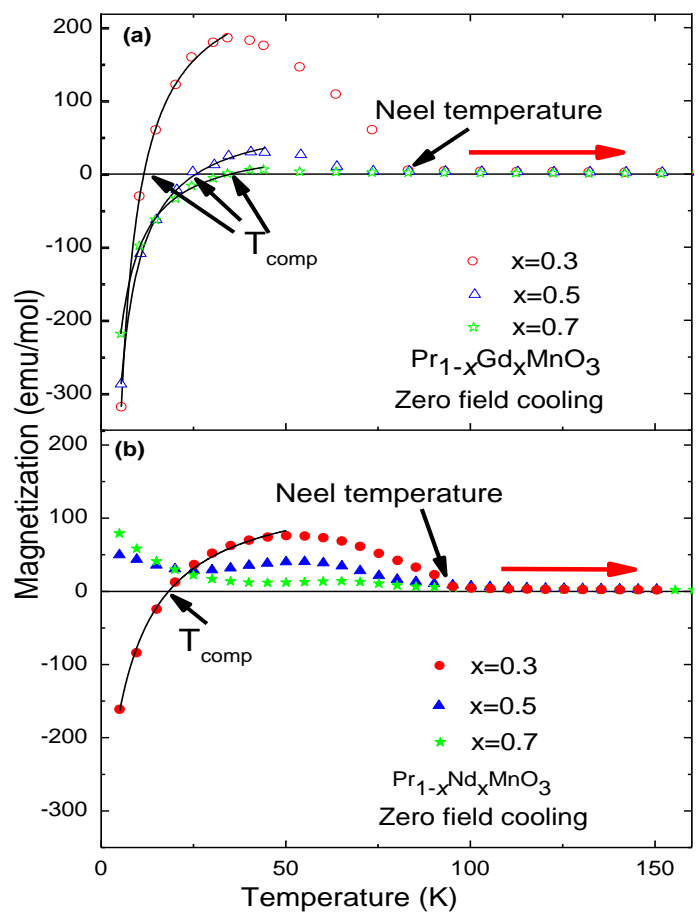

**Figure 2**

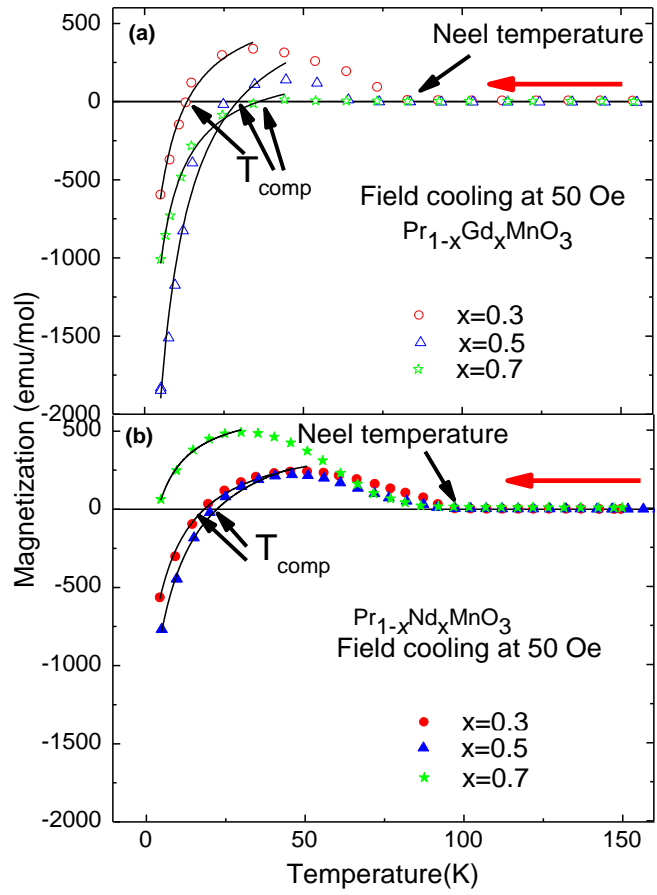

**Figure 3**

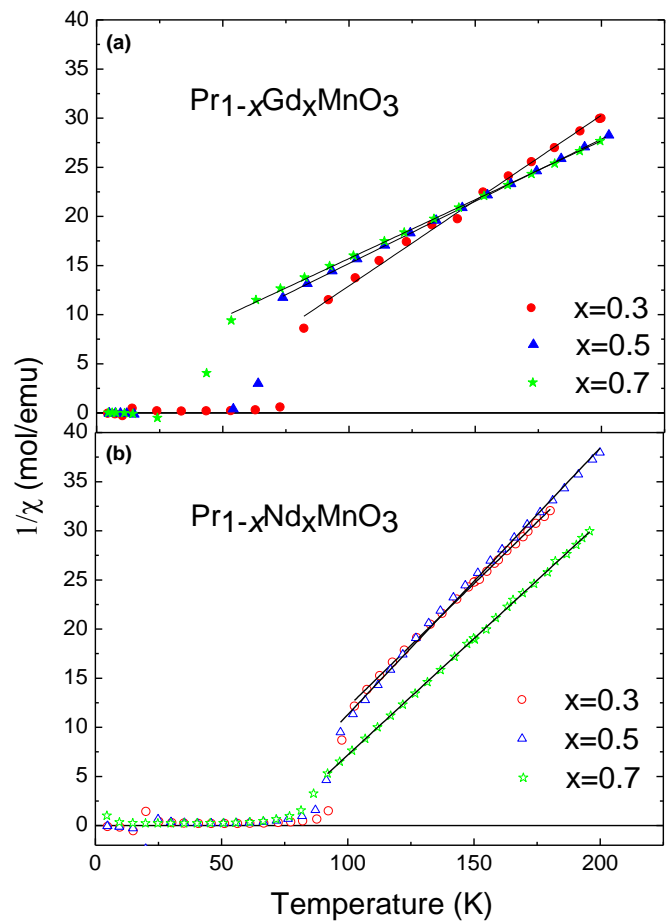

**Figure 4**

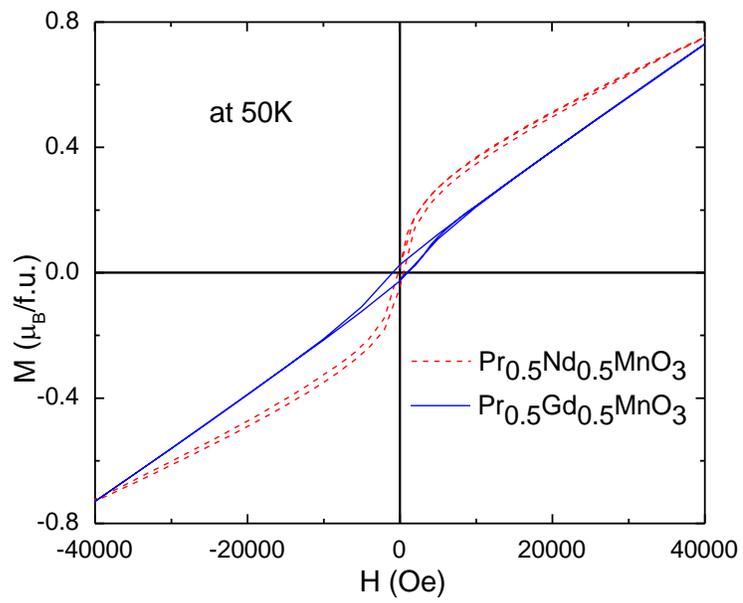

**Figure 5**

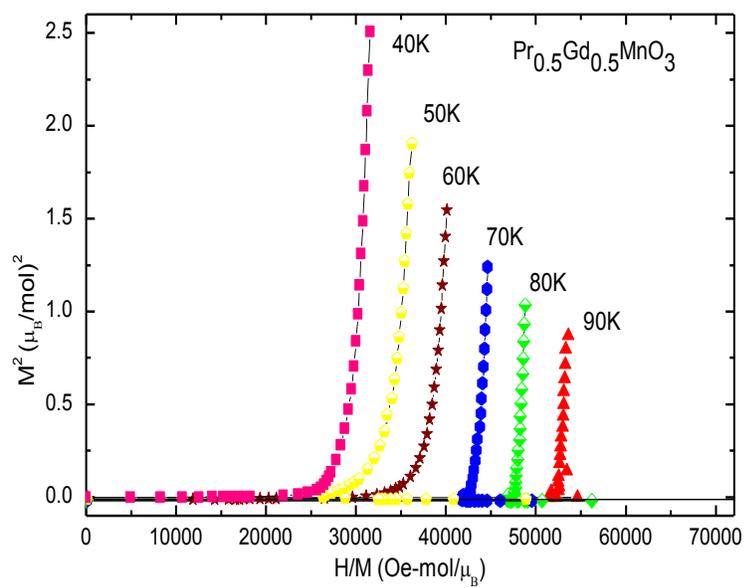

**Figure 6**